\documentclass[traditabstract]{aa} 
\usepackage{graphicx}
\usepackage{txfonts}
\usepackage{longtable}
\usepackage{natbib}
\usepackage[breaklinks=true]{hyperref} 

\def\a{{$\alpha$}}

\newcommand{\h}{$^{\rm h}$}
\newcommand{\m}{$^{\rm m}$}
\newcommand{\s}{$^{\rm s}$}
\newcommand{\dd}{$\delta$}
\newcommand{\ha}{\rm H$\alpha$}
\newcommand{\hbeta}{\rm H$\beta$}
\newcommand{\HII}{\ion{H}{ii}}
\newcommand{\hnii}{{\rm H}$\alpha+[$\ion{N}{ii}$]$}
\newcommand{\nii}{$[$\ion{N}{ii}$]$}
\newcommand{\sii}{$[$\ion{S}{ii}$]$}
\newcommand{\oi}{$[$\ion{O}{i}$]$}

\newcommand{\oiii}{$[$\ion{O}{iii}$]$}

\newcommand{\flux}{$10^{-17}$ erg s$^{-1}$ cm$^{-2}$ arcsec$^{-2}$}

\newcommand{\dens}{\rm cm$^{-3}$}
\newcommand{\vel}{\rm km s$^{-1}$}
\newcommand{\sulfur}{[S~{\sc ii}]}
\newcommand{\nitrogen}{[N~{\sc ii}]}
\newcommand{\oxygen}{[O~{\sc iii}]}
\newcommand{\siirat}{$[$\ion{S}{ii}$]\lambda\lambda\ 6716/6731$} 
\newcommand{\HNII}{{\rm H}$\alpha+$[N {\sc ii}]~6548~\&~6584~\AA}  
\newcommand{\OIII}{$[$\ion{O}{iii}$]$~5007~\AA}
\newcommand{\SII}{$[$\ion{S}{ii}$]$~6716~\&~6731~\AA}
\begin{document}
   \title{Discovery of optical candidate supernova remnants in Sagittarius}


   \author{J. Alikakos
          \inst{1,2}
          \and
          P. Boumis\inst{1}
          \and
          P. E. Christopoulou\inst{2}
          \and
          C. D. Goudis\inst{1,2}
          }

   \institute{Institute of Astronomy \& Astrophysics\thanks{Renamed to: Institute of Astronomy, Astrophysics, Space Applications and Remote Sensing}, National
Observatory of Athens, I. Metaxa \& V. Pavlou, P. Penteli, GR-15236
Athens, Greece\\
              \email{[johnal;ptb;cgoudis]@astro.noa.gr}
         \and
             Astronomical Laboratory, Department of Physics, University of
Patras, GR-26500 Rio-Patras, Greece\\
             \email{pechris@upatras.physics.gr}
             }

   \date{Received; accepted}

 
  \abstract
   {During an [O {\sc iii}] survey for planetary nebulae, we identified a region in Sagittarius containing several candidate Supernova Remants and obtained deep optical narrow-band images and spectra to explore their nature. The images of the unstudied area have been
obtained in the light of \hnii, \sii\ and \oiii. The resulting mosaic
covers an area of $1.4\degr \times 1.0\degr$~where filamentary and
diffuse emission was discovered, suggesting the existence of more than
one supernova remnants (SNRs) in the area. Deep long slit spectra were
also taken of eight different regions. Both the flux calibrated images
and the spectra show that the emission from the filamentary structures
originates from shock-heated gas, while the photo-ionization mechanism is
responsible for the diffuse emission. Part of the optical emission is
found to be correlated with the radio at 4850 MHz  suggesting their
association, while the WISE infrared emission found in the area at 12 and 22 $\mu$m marginally correlates with the optical. The presence of the \oiii\ emission line in one of the
candidate SNRs suggests shock velocities into the interstellar
"clouds" between 120 and 200 km s$^{-1}$, while the absence in the
other indicates slower shock velocities. For all candidate remnants
the [S {\sc ii}] $\lambda\lambda$ 6716/6731 ratio indicates electron
densities below 240 cm$^{-3}$, while the \ha\ emission has been
measured to be between 0.6 to 41$\times$\flux. The existence of eight
pulsars within 1.5$\degr$ away from the center of the candidate SNRs
also supports the scenario of many SNRs in the area as well as that the detected optical emission could be part of a number of supernovae explosions.}


  \maketitle
%

\section{Introduction}
Supernova explosions belong to the most spectacular events in
the Universe. Observations of galaxies reveal several events every year \citep{manu05},
where the supernova is of comparable brightness to the entire galaxy
for days up to weeks. Supernova remnants (SNRs) which are the
consequent result of such events are some of the strongest radio sources
observed. SNRs have a major influence on the properties of the
interstellar medium (ISM) and on the evolution of galaxies as a
whole. They enrich the ISM with heavy elements, release about 10$^{51}$
ergs and heat the ISM, compress the magnetic field and efficiently accelerate in
their shock waves energetic cosmic rays as observed
throughout the Galaxy. The majority of known SNRs have been discovered
by their non-thermal radio emission \citep{gre09} while a
smaller number of them are observed in other wavelengths (e.g optical;
\citealt{bou08, bou09}, X-rays; \citealt{rey09}, infrared; \citealt{rea06}).

In this paper, we report the optical detection of many filamentary and
diffuse structures (possibly more than one SNR) in the region of
Sagittarius constellation. During an \OIII\ survey for planetary nebulae \citep{bou03,bou06}, we identified a very strong \sii\ source designated as candidate supernova remnant instead of
planetary nebula. Following that detection, a number of images in
\hnii, \sii\ and \oiii\ were taken in order to explore the SNR candidate
area and many filamentary structures were discovered. Only one known
SNR was found in the area (G 16.2--2.7; \citealt{tru99}), hence
all the other filamentary structures were examined in detail in order
to identify their origin. Information about the observation and data
reduction is given in Sect. 2. In Sect. 3 and 4 the results of the
imaging and spectra observations are presented, while in Sect. 5 we
report on observations in wavelengths other than the optical. Finally, in
Sect. 6 we discuss the properties of the new candidate SNRs.

\section{Observations}
A summary and log of our observations are given in
Table~\ref{table1}. In the sections below, we describe these
observations in detail.

\subsection{Imaging}

All images were taken with the 0.3 m Schmidt--Cassegrain (f/3.2)
telescope at Skinakas Observatory in Crete, Greece, in 2005 June 7, 8
and 9, and August 26 and 28. A 1024 $\times$ 1024 Thomson CCD was used
which has a pixel size of 19 $\mu$m resulting in a 70\arcmin\ $\times$
70\arcmin\ field of view and an image scale of 4\arcsec\
pixel$^{-1}$. The area of interest was observed for 2400 s in \hnii,
\sii\ and \oiii\ filters, while corresponding continuum images were
also observed (180 s each) and after the appropriate scaling were
subtracted from the narrow-band images to eliminate the
confusing star field. The continuum subtracted images of the \hnii\
and \sii\ emission lines are shown in Figs. \ref{fig1} and \ref{fig2}.

The image reduction was carried out using the IRAF and MIDAS
packages. All frames were bias subtracted and flat--field corrected
using a series of twilight flat--fields. The absolute flux calibration
was performed through observations of a series of spectrophotometric
standard stars (HR5501, HR7596, HR7950, HR8634 and HR9087; \citealt{ham92}). The astrometric solution for all data frames was
calculated using the Space Telescope Science Institute (STScI) Guide
Star Catalogue II (GSC--II; \citealt{las08}). All the
equatorial coordinates quoted in this work, refer to epoch 2000.

In order to cover in full the area of interest, we used wide--field
images from the SuperCOSMOS \ha\ Survey (SHS; \citealt{par05}). The resulting 1.4\degr $\times$ 1.0\degr\ mosaic was
created from 16 different fields, each one having a 30\arcmin\ $\times$
30\arcmin\ field of view and an image scale of 0.67\arcsec\
pixel$^{-1}$. The mosaic was used in order to compare the detected
optical emission with other wavebands (radio and infrared -- see
Sect. 3.3 and Figs.~\ref{fig5}, \ref{fig6}). The details of all
imaging observations are given in Table \ref{table1}.

\subsection{Spectroscopy}
Low dispersion long--slit spectra were obtained with the 1.3 m
Ritchey--Cretien (f/7.7) telescope at Skinakas Observatory in 2005
June 4 and 5, July 10 and September 6 and 7. The 1300 line mm $^{-1}$
~grating was used together with a 2000 $\times$ 800 SITe CCD (15
$\times$ 15 $\mu$m$^2$~pixels) resulting in a scale of 1\AA\
pixel$^{-1}$ and covering the range 4750 \AA\ -- 6815 \AA. The above
combination results in a spectral resolution of $\sim$8 and $\sim$11
\AA\ in the red and blue wavelengths, respectively.  The slit width
was 7\farcs7 and its length 7\farcm9 and in all cases was oriented in
the south--north direction. The spectrophotometric standard stars
HR4468, HR5501, HR7596, HR9087, HR8634, and HR7950 \citep{ham92} were
observed to calibrate the spectra. The data reduction was performed
using the IRAF package (twodspec).

The deep low resolution spectra were taken on the relatively bright
optical filament (their exact positions are given in
Table~\ref{table1}). In Table~\ref{table4}, we present the relative
line fluxes taken from different apertures (a, b and c) along each
slit. In particular, apertures a, b and c have an offset (see
Table~\ref{table1}) north or south of the slit center which were
selected because they are free of field stars in an otherwise crowded
field and they include sufficient line emission to permit an accurate
determination of the observed line fluxes (their exact aperture length
is given in Table~\ref{table1}). The background extraction aperture
was taken towards the northern end or the southern end of the slits
depending on the slit position. The signal to noise ratios presented
in Table~\ref{table4}~do not include calibration errors, which are
less than 10 percent. Typical spectra are shown in Fig. \ref{fig3}.

%
%
\section{Results}

\subsection{The optical emission line images}

\begin{figure*}
\centering
\includegraphics[scale=0.75]{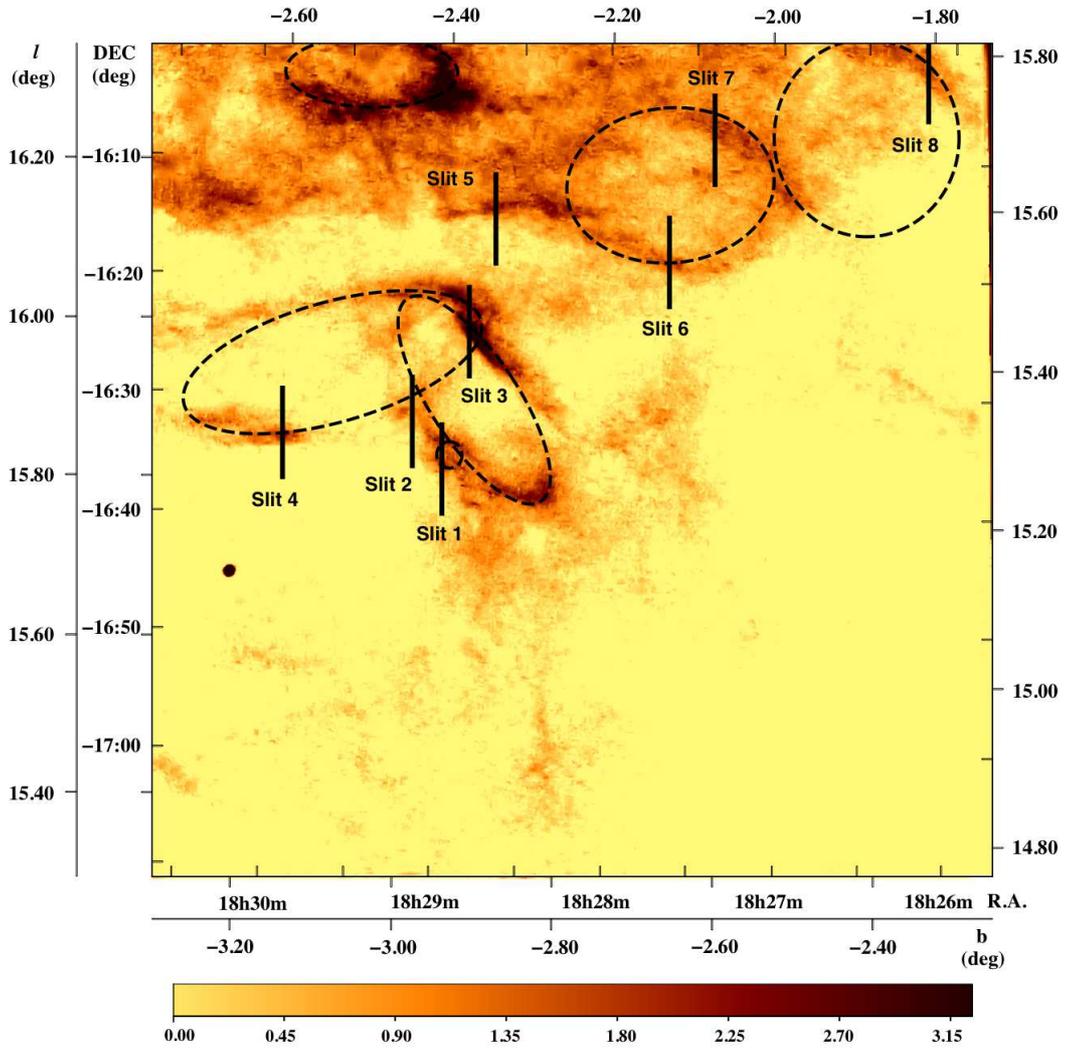}
\caption{The continuum--subtracted image of the observed area in Sagittarius, in
\hnii\ emission in both equatorial (R.A., Dec internal lines on axis) and galactic (l, b external lines on axis) coordinates. The black lines indicate the positions of the slits and the dashed-ellipses the geometry  of the newly discovered candidate SNRs. Shadings run linearly from 0 to 3.2$\times$\flux.}
\label{fig1}
\end{figure*}

\begin{figure*}
\centering
\includegraphics[angle=-90, scale=0.70]{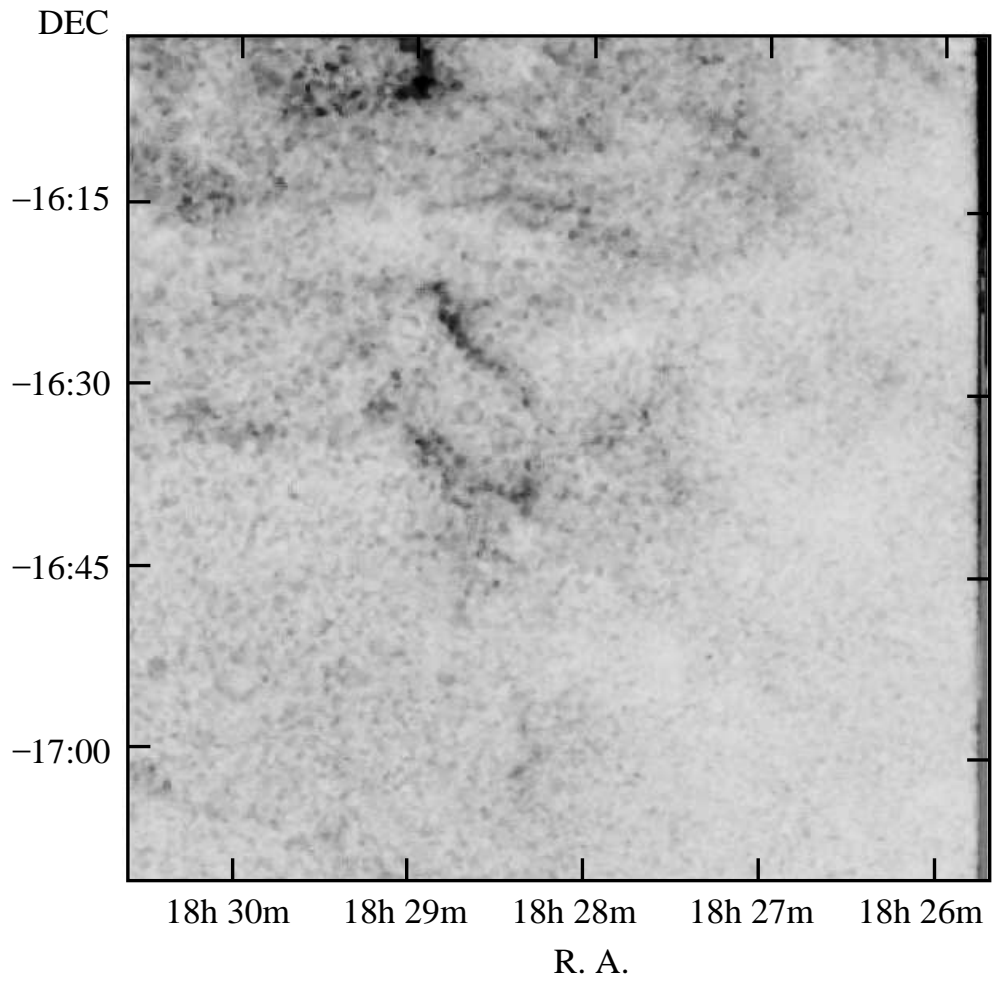}
\caption{The continuum--subtracted image of the observed area in
\sii\ emission.  Shadings run linearly from 0 to 90$\times$\flux.}
\label{fig2}
\end{figure*}

\begin{figure*}
\centering
\includegraphics[width=\textwidth]{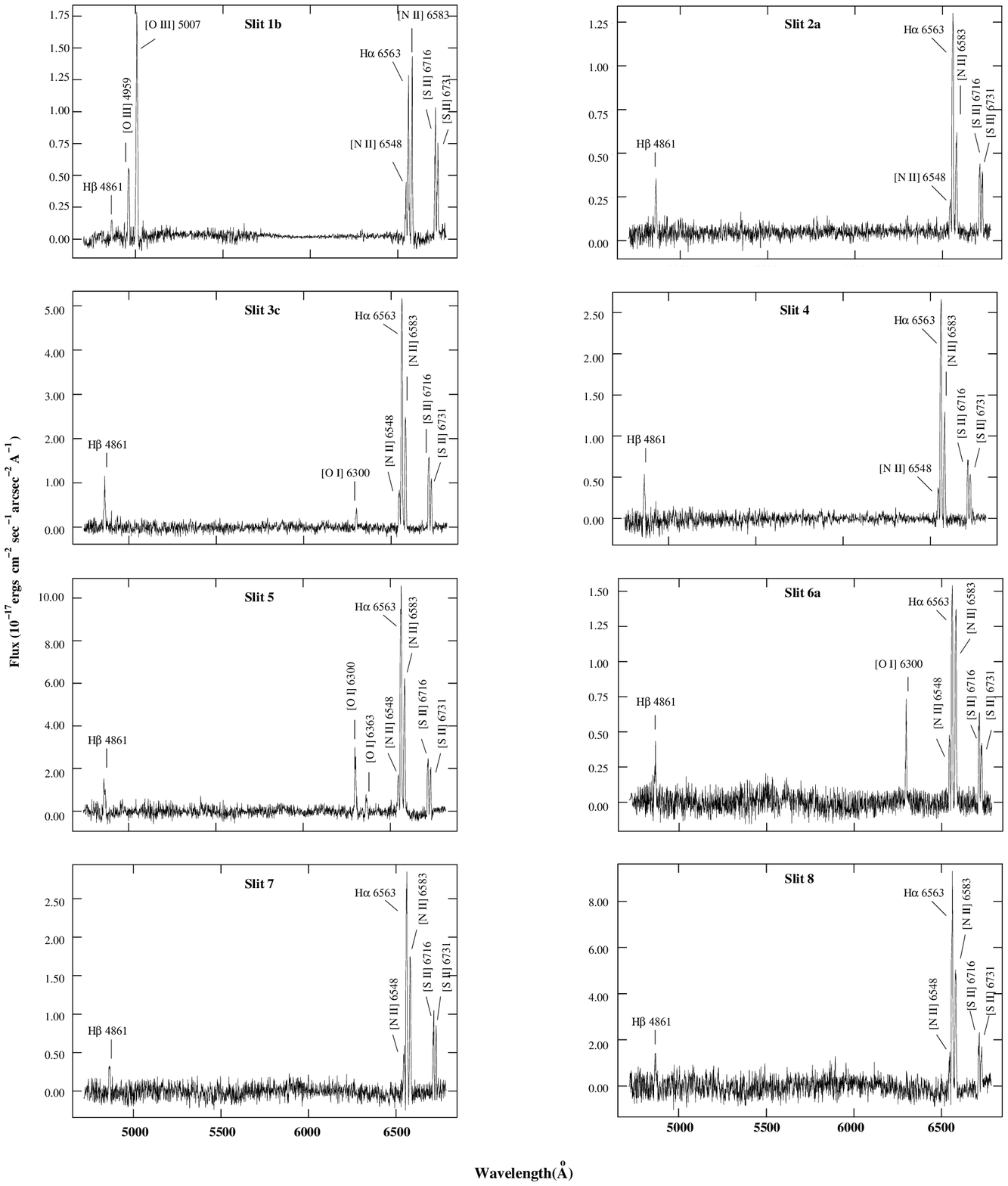}
\caption{Long--slit spectra from different positions of the observed
area (see Table \ref{table1}).}
\label{fig3}
\end{figure*}

The images in Figs. \ref{fig1} and \ref{fig2} show considerable new faint optical
emission including filamentary and diffuse structures. The \hnii\
(Fig. \ref{fig1}) best describes the newly detected structures,
while the \sii\ image (Fig. \ref{fig2}) also shows strong emission
but less filamentary structure. In contrast, \oiii\ emission was only detected
in one small area and it is not shown here.

All images being flux calibrated provide a first indication of the
nature of the observed emission (see Table~\ref{table2}). A study of
these images shows that most parts of the optical emission originate
from shock heated gas since we estimate ratios \sii/\ha $\geq$0.47.
This conclusion is verified by the deep long--slit spectra which offer
more accurate measurements of the individual line fluxes. The variety
of structures detected in the \hnii\ and \sii\ images are not present
in the medium ionization line of \oiii\ apart from one area (Slit 1),
hence the 3$\sigma$~upper limit, over the area is given in Table
\ref{table2}.

Several thin and curved filaments can be seen in Fig. \ref{fig1}. In
general, the field appears somewhat complex due to the presence of
many filamentary structures and significant number of diffuse emission
structures. A search in the database did not reveal any known bright
optical nebula. A combination of their morphology and the
spectroscopic results suggest the existence of more than one possible
SNR in the region. In particular, following their geometry we suggest
the existence of six candidate SNRs (indicated as dashed-ellipses in
Fig. \ref{fig1}). Their names, centers and diameters are presented in
Table \ref{table3}. Of course, the possibility of less or more SNRs in
the region cannot be ruled out.

The most interesting filaments of the candidate SNRs are described
here in more detail.

{\it G 15.6$-$2.6}. The basic characteristic of its filaments is their
brightness in \hnii\ (Fig. \ref{fig1}) and \sii\
(Fig. \ref{fig2}). The bulk of the emission seems to be bounded by at
least two very bright complex filaments. The southern structure covers
the area from \dd$\sim-$16\degr39\arcmin, \a$\simeq$18\h28\m21\s\ to
\dd$\sim-$16\degr33\arcmin, \a$\simeq$18\h29\m01\s, while the
northern, having the same inclination ($\sim$45\degr), with respect to
the east--west direction, covers the area between
\dd$\sim-$16\degr29\arcmin, \a$\simeq$18\h28\m31\s\ and
\dd$\sim-$16\degr21\arcmin, \a$\simeq$18\h28\m55\s. No significant
emission was found in the image of the \oiii\ medium ionization
line. The morphology of the \sii\ image is generally similar to,
though not as bright as, that of the \hnii\ image. We detected \sii\
emission where most of the \hnii\ emission was found with filamentary
bright structures in the south--east and north--west areas, while
diffuse emission characterizes the rest of the remnant's
emission. Diffuse emission and several shorter filamentary structures
are detected between the north and south boundaries. The images show
\sii/\ha $\geq$0.7 in agreement with the two spectra positions (slits
1 \& 3) which show \sii/\ha $\geq$1.0. It should be noted that part of
the filaments in the south and to the north seem to correlate with the radio and infrared emission but their resolution does not permit to verify or not such a correlation (see Section 3.3). New pointed radio observations are needed to come into secure conclusion.

{\it G 15.8$-$2.8}. The most interesting regions lie in the
south--east and north--west, where bright filamentary structures are
present (between $\alpha$$\simeq$18\h29\m45\s, $\delta
$$\simeq$--16\degr34\arcmin20\arcsec; $\alpha$$\simeq$18\h30\m26\s,
$\delta$$\simeq$--16\degr32\arcmin17\arcsec\ and $\alpha$$\simeq$18\h28\m48\s,
$\delta$$\simeq$--16\degr22\arcmin45\arcsec; $\alpha$$\simeq$18\h28\m58\s,
$\delta$$\simeq$--16\degr21\arcmin15\arcsec). There are also fainter
filaments and diffuse emission which cover most of the suggested SNR
area, both to the north (from the bright north--west filament to
$\alpha$$\simeq$18\h30\m33\s, $\delta$$\simeq$--16\degr25\arcmin00\arcsec) 
and south (from the bright south--east filament to
$\alpha$$\simeq$18\h28\m47\s, $\delta$$\simeq$--16\degr25\arcmin30\arcsec).
Similarly to the previous candidate SNR, no \oiii\ emission was detected.
Both images and spectra (slits 2 \& 4) show similar \sii/\ha\ (between
0.47 and 0.60). Part of this SNR to the west overlaps with the north
filament of the previous SNR, however their morphology and curvature
suggest that the filamentary structures are separated and they
probably do not correlate with each other. Also, there is a gap of
$\sim$1\arcmin\ between the very bright filaments and they both show
infrared emission, while it seems that only the filament from this candidate SNR correlates with the radio emission (Section 3.3). However, due to the low resolution of the radio observations, the possibility that the radio emission comes from both SNRs cannot be ruled out. It should be noted that because the radio emission is generally weak, the overlap of the two remannts would certaintly make it brighter. Also, the fact that the high frequency radio emission correlates better with the optical than the low frequency, which might suggests a flatter radio spectrum and therefore thermal radio emission, should be examined further by new pointed radio observations in order to verify their nature.

{\it G 15.8$-$2.2}. The faint filamentary and diffuse emission of this
candidate SNR forms a well defined ellipse (Fig. \ref{fig1}). There is
a bright structure to the east ($\alpha$$\simeq$18\h28\m20\s,
$\delta$$\simeq$--16\degr14\arcmin25\arcsec) and many small filaments
all around the SNR's borders. The \sii/\ha\ emission suggests
shock--heated mechanisms ($\sim0.6$) in agreement with that from the
spectra (slits 6 \& 7) which is between 0.45 and 0.7. There is also a
bright filament further to the east (from
$\alpha$$\simeq$18\h28\m28\s, $\delta$$\simeq$--16\degr14\arcmin00\arcsec\ to
$\alpha$$\simeq$18\h28\m56\s, $\delta$$\simeq$--16\degr15\arcmin05\arcsec)
with shock--heated emission, having however the weakest  \sii\  measured in the area (\sii/ha$=$0.45; slit 5). Following its morphology, it is
probable that it is not related to this remnant, however, their
correlation cannot be ruled out completely. There is also radio and
infrared emission which might correlates with the optical in some of the
SNR's regions (Section 3.3), while \oiii\ emission has not been
detected.

{\it G 15.8$-$1.9}. In contrast to the previous SNR candidates, this
is the fainter one in both \hnii\ and \sii\ images, having less
filamentary and more diffuse emission, while \oiii\ was not
detected. Its brightest part is to the north--northwest (from
$\alpha$$\simeq$18\h26\m00\s, $\delta$$\simeq$--16\degr04\arcmin25\arcsec\ to
$\alpha$$\simeq$18\h26\m40\s, $\delta$$\simeq$--16\degr00\arcmin20\arcsec), 
there is fainter emission to the east ($\alpha$$\simeq$18\h26\m50\s,
$\delta$$\simeq$--16\degr13\arcmin45\arcsec), while no emission has
been detected to the south. The \sii/\ha\ ratios of 0.48 (spectrum,
slit 8) and 0.50 (imaging) suggest that a shock--heated mechanism
produced the emission while the optical emission correlates well with
the radio and partially with the infrared (Section 3.3).
 
{\it G 16.2$-$2.5}. This is the brightest candidate in \hnii\ and
\sii\ with \sii/\ha $\sim 0.5$. There are bright thick filaments
almost all around the remnant (south from $\alpha \simeq$
18\h28\m45\s, $\delta \simeq$ --16\degr06\arcmin10\arcsec\ to $\alpha
\simeq$ 18\h29\m55\s, $\delta \simeq$ --16\degr03\arcmin30\arcsec;
west from $\alpha \simeq$ 18\h28\m45\s, $\delta \simeq$
--16\degr06\arcmin40\arcsec\ to $\alpha \simeq$ 18\h28\m56\s, $\delta
\simeq$ --16\degr01\arcmin30\arcsec\ and north--west to $\alpha
\simeq$ 18\h29\m14\s, $\delta \simeq$ --15\degr59\arcmin45\arcsec)
with a width of $\sim$2--3\arcmin (due to the position of this
candidate SNR, the north part can only be seen in
Fig. \ref{fig5}). Weaker and diffuse emission patches are located in
its north--east area while no \oiii\ emission was found. The detected
emission line structures are also bounded in all directions by diffuse
emission which is probably due to \HII\ regions. It is very close to
the only known radio SNR G 16.2--2.7 but their position and morphology
suggest that they are not related to each other (see
Fig. \ref{fig5}). However, a small possibility that this emission is
related to the radio remnant still remains if we consider that they
can be seen by a different angle to the line of sight. Kinematic
oobservations could confirm or reject the latter scenario. There is
also infrared emission in the area which partially correlates with the
optical (Section 3.3).

{\it G 15.6$-$2.7}. This is the most peculiar region since it is the
only area that shows strong \oiii\ and very strong \nii\ emission. A
shock--heated mechanism is responsible for the emission found since
the \sii/\ha $>$1.0. Its morphology, size and position in the diagnostic diagram (Fig. 4) let us conclude that it
should be a SNR which might not related to G 15.6$-$2.6, however their
correlation cannot be ruled out so further investigation (i.e. high resolution spectroscopic observations) is needed in order to identify its origin. In case they are related, then the possibility that its \sii\ emission is background emission from G 15.6$-$2.6 and it does not come from this object will change the responsible mechanism to photoionization and then it might not be an SNR but a planetary nebula. Note that the strong \oiii\ emission is not unusual to be found in both SNRs and PNe so it cannot be used as a distiguishable criterion. In Table~\ref{table2} a typical
\oiii\ flux is given. Radio emission is not found to be correlated
with this SNR candidate while there is faint infrared emission in its vicinity which might associate with it. 
(Section 3.3).

\subsection{The optical spectra}
Deep long--slit spectra were taken in order to accurately determine
the nature of the observed emission by measuring the strengths of the
\ha\ and \sii\ emission lines. Eight different spectra, extracted from
the relatively brightest optical filaments, are shown in
Fig. \ref{fig3} and the measured fluxes are given in Table
\ref{table4}. Different apertures (a, b and c) were extracted from
each spectrum along the slit which have an offset (see
Table~\ref{table1}) north or south of the slit center. The selection
criteria were to be in an area which is free of field stars and they
include sufficient line emission to permit an accurate determination
of the observed line fluxes. The background extraction aperture was
taken towards the northern end or the southern end of the slits
depending on the slit position. The signal to noise ratios presented
in Table~\ref{table4}~do not include calibration errors, which are
less than 10 percent.

The properties of the spectra strongly point to emission from shock
heated gas (\sii/\ha\ ratios between 0.44 to 1.20, \nii/\ha\ ratios
between 0.53 to 1.35; presence of \oi\ emission; Table
\ref{table4}). The filamentary nature of the newly discovered optical
radiation, as seen in the narrow band images, also supports this
conclusion. The absolute \ha\ flux covers a range of values from 0.6
to 41.3 $\times$ \flux. The \siirat\ ratio that was measured between
1.2 and 1.5, indicates low electron densities (below 240 cm$^{-3}$;
\citealt{ost06} and below 400 cm$^{-3}$~by taking
into account the statistical errors on the sulfur lines; \citealt{sha95}).

Although, the low ionization lines are quite strong, \oiii\ line
emission at 5007 \AA\ is only detected at pos. 1 (Table \ref{table2}),
suggesting shock velocities $\sim$120 km s$^{-1}$ \citep{cox85} and below 200 km s$^{-1}$ \citep{all08} in that area. Also, the \oiii/\hbeta\ is $>$ 6,
according to theoretical models \citep{ray88}, suggest
the presence of shocks with incomplete recombination zones. The
relatively weak \hbeta\ emission and the absence of \oiii\ emission in
all the other positions suggest significant interstellar attenuation
of the optical emission and they could be explained by slow shocks
propagating into the interstellar clouds ($\leq$ 100 \vel; \citealt{har87}).

The log(\ha/\nii) versus log(\ha/\sii) intensities, from
Table~\ref{table4}, corrected for interstellar extinction, are
compared with others of well defined phenomena in Fig. \ref{fig4}
(following \citealt{sab77} \& \citealt{can81})
and show that all positions are within or very close to (taking into
account the calculated error for those which are not within) the area
designated for SNRs.

\subsection{Observations at other wavelengths}
Radio emission in the area of the optical structures is detected in the low
resolution (7\arcmin) 4850 MHz images of the Green Bank survey
\citep{con91}. In most of the cases, the radio emission matches with the optical emission but given the low resolution, it is
hard to make a detailed spatial correlation apart from confirming their association. We have also examined the
higher resolution CGPS data at 1420 MHz \citep{tay03}
 but no prominent emission was detected to be correlated with the optical. The observed filaments
are located close to radio contours (Fig.~\ref{fig5}), in many cases there is a good correlation but the low resolution of the
radio images does not allow us to determine the relative position of
the filament with respect to the shock front. Higher resolution radio observations in different wavelengths should be performed to provide conclusive evidence on the nature of the sources as SNRs.  

We have also searched ROSAT All--sky survey data for X--ray emission 
but it is very faint and diffuse and we can not identify the sources
in detail.
We also searched the VLA Galactic Plane Survey (VGPS; \citealt{sti06}) for H{\sc i} kinematics data but unfortunately the area of interest is not covered by this survey.

 In order to explore how the
optical emission correlates with the infrared, we searched all the available data at NASA/IPAC Infrared Science Archive. In particular, we examined the GLIMPSE/Spitzer \citep{chu09}, the  WISE \citep{wri10} and the IRIS \citep{miv05} images of the same area. Unfortunately, the area we are interested in was not covered by the GLIMPSE survey at 3.6, 4.5, 5.8 and 8.0 $\mu$m (resolution at 1.2\arcsec) while the IRIS at 25, 60 and 100 $\mu$m (resolution from 30\arcsec\ to 2\arcmin) data revealed clear enhancement of infrared emission in the area where the optical emission of the candidate SNRs is detected, however the low-resolution maps do not permit a detailed comparison.  On the other hand, the WISE data at 3.4, 4.6, 12 and 22 $\mu$m (resolution of 6.1, 6.4, 6.5 and 12\arcsec, respectively) show evidence of correlation with the optical and/or radio emission in the last two channels.
In particular, both the 12 and 22 $\mu$m  show evidence of association with the optical and radio filamentary structures of the candidate SNRs. Since the 12 $\mu$m image has essentially the same features with the 22 $\mu$m but they appear to be stronger in the latter, in Fig.~\ref{fig6} we present a greyscale representation of the 22 $\mu$m image with
overlapping contours of the optical emission. 
 In a few cases, the infrared emission seems to follow the morphology of the candidate 
SNRs suggesting their possible association which is expected since it is known that the shocked gas cools through emission lines, and many important emission lines occur in the mid--infrared. A similar case is that of the SNR G 296.7$-$0.9 \citep{rob12} where possible association of this SNR with the infrared emission was found. It should also be noted that the appearance of the infrred emission in the vicinity and not on the optical emission in most of the cases is expected, since in general there is not a very good correlation between the infrared and the  H$\alpha$ emission  \citep{rea06}.

%

\section{Discussion}

The newly discovered candidate SNRs towards the Sagittarius  constellation show up as incomplete circular or elliptical structures in the optical and in most of the cases in the radio and without any X--ray and H {\sc i} emission detected so far. The absence of soft X--ray emission may indicate a low shock temperature and/or a low density of the local interstellar medium.
Its optical emission marginally correlates with the infrared. 
The elliptical shape of the candidate SNRs is unusual compared to most known SNRs suggesting that the surrounding medium is very irregular with not constant interstellar density. However, it should be mentioned that a significant number of known SNRs show not circular structure (bilateral, barriel, elliptical, cilindrical etc.) depending on their position to the line of sight so this might also be a reason of their appearance.

Detailed optical observations have been performed in an attempt to
understand the nature of the candidate SNRs. The lower ionization images in
\hnii\ and \sii\ reveal
several filamentary and diffuse structures while the higher ionization image in \oiii\  shows emission only in one region. 
The \hnii\ image best describes the newly detected
structures. Sulfur line emission is also detected and generally
appears less filamentary and more diffuse than in the \hnii\ image
with their position and shape in agreement with that of the \hnii. The \oiii\ flux
production depends mainly on the shock velocity and the ionization
state of the preshocked gas. Therefore, as mentioned in Sect. 3.1, the
absence of \oiii\ emission in almost all of the areas may be explained by slow shocks propagating into the ISM. 
The presence of [O {\sc i}] 6300 \AA\ line emission is also consistent with the emission being from shocked material.
The \oiii/\hbeta\
ratio is a very useful diagnostic tool for complete or incomplete
shock structures \citep{ray88}. However, the absence of \oiii\ does not allow us to suggest for complete or incomplete shock structures apart from pos.1 where shocks with incomplete recombination zones should be present.
Both the calibrated images and the long--slit
spectra suggest that the detected emission results from shock heated
gas since the \sii/\ha\ ratio exceeds the empirical SNR criterion
value of 0.4--0.5, while the measured \nii/\ha\ ratio also confirms
this result.
\par
\begin{figure*}
\centering
\includegraphics[scale=0.75]{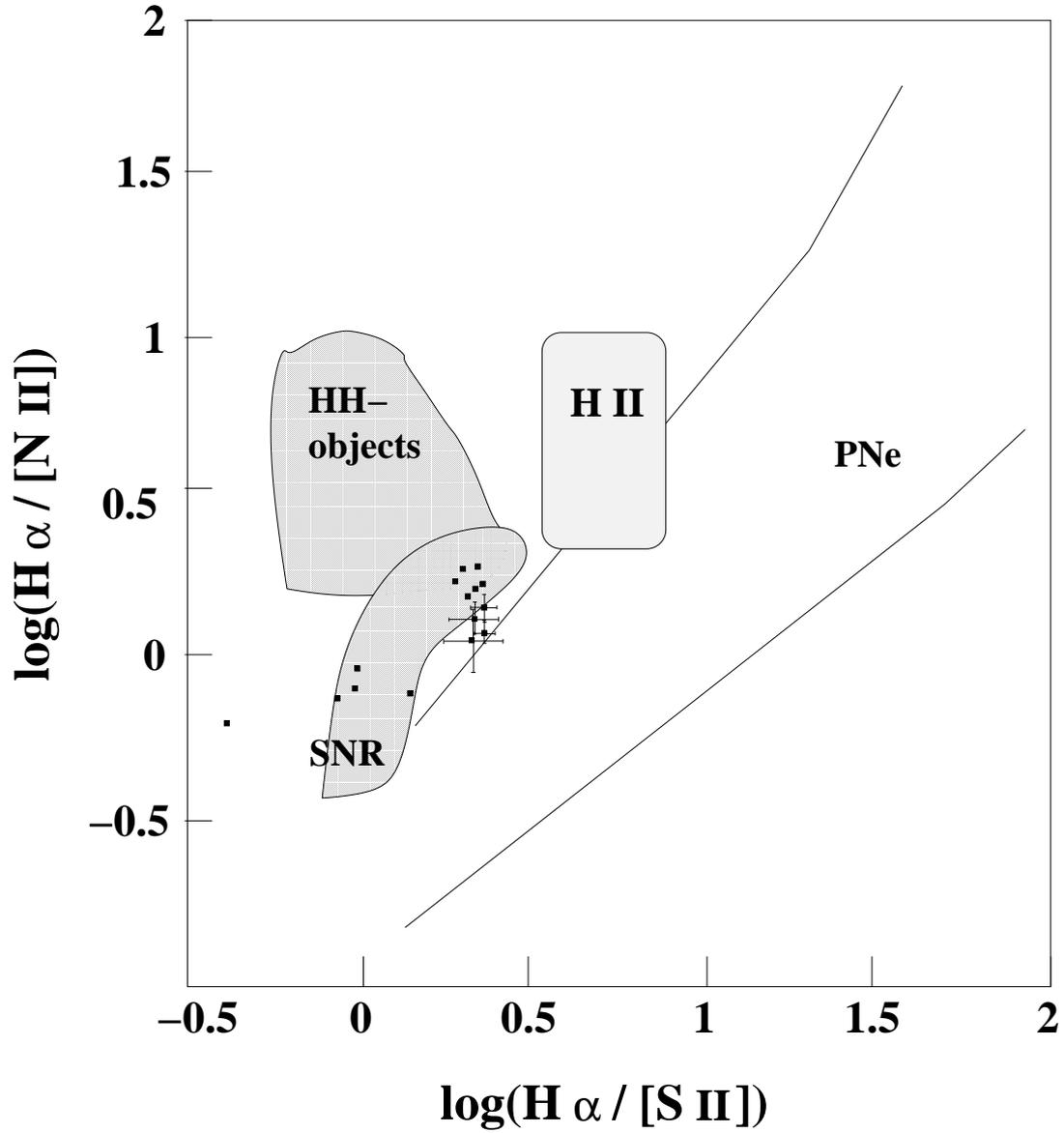}
\caption{Diagnostic diagram (\citet{sab77}; \citet{can81}), where the positions of line ratios listed in Table \ref{table1},
from Slit 1a to 8, are shown in black squares.  For those which are not within the SNR region the calculated errors have taken into account.}
\label{fig4}
\end{figure*}
\begin{figure*}
\centering
\includegraphics[width=\textwidth]{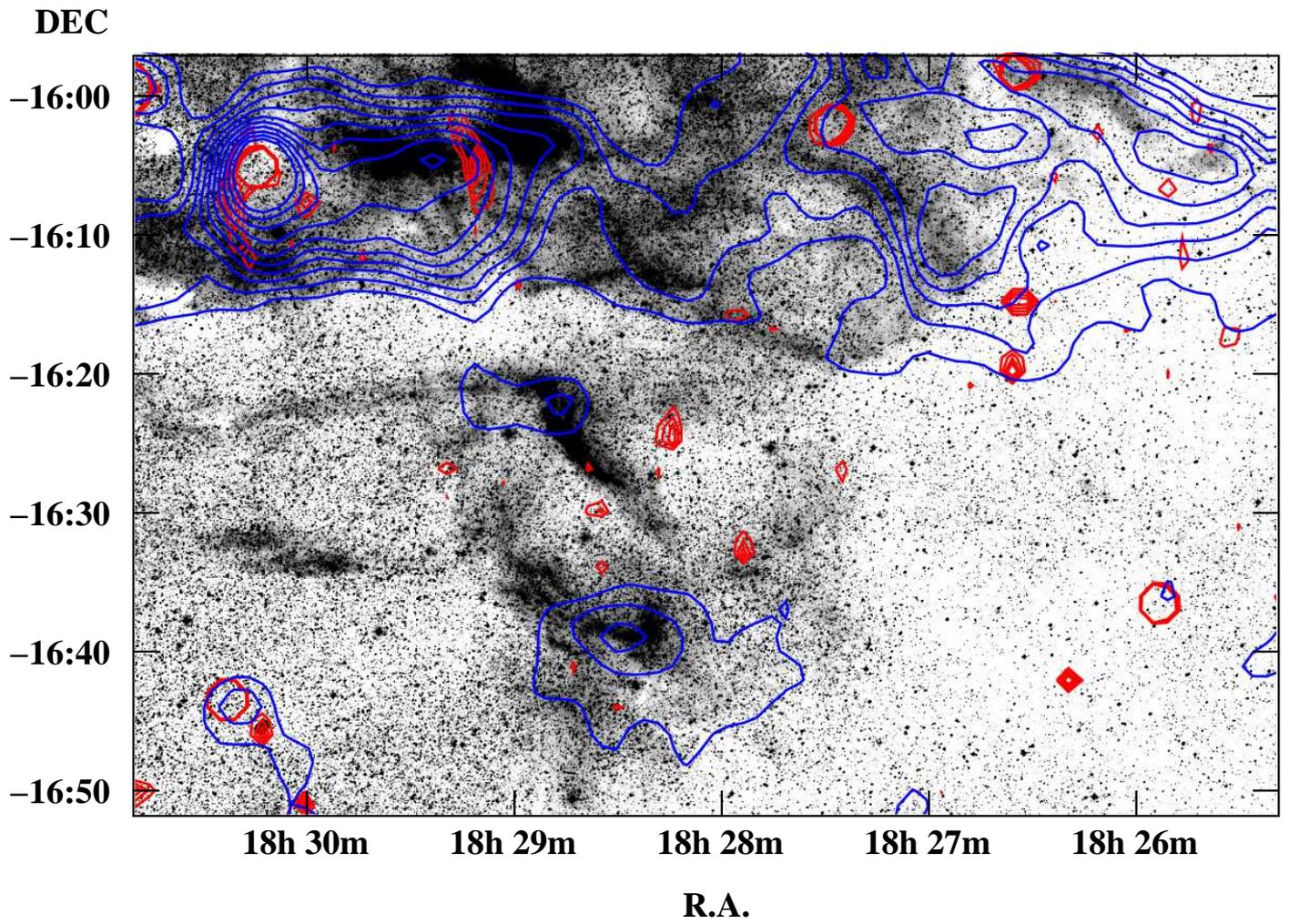}
\caption{The correlation between the SuperCOSMOS \ha\ mosaic image and the radio emission
from archival data at 4850 MHz (blue line) and 1400 MHz (red line). The blue contours
scale linearly from 2.0$\times 10^{-2}$~Jy/beam to 0.2 Jy/beam, with
step 0.02 Jy/beam and the red from 1.2$\times 10^{-3}$~Jy/beam to
4.0$\times 10^{-3}$~Jy/beam, with step 9.3$\times 10^{-4}$ Jy/beam.}
\label{fig5}
\end{figure*}
\begin{figure*}
\centering
\includegraphics[scale=0.75]{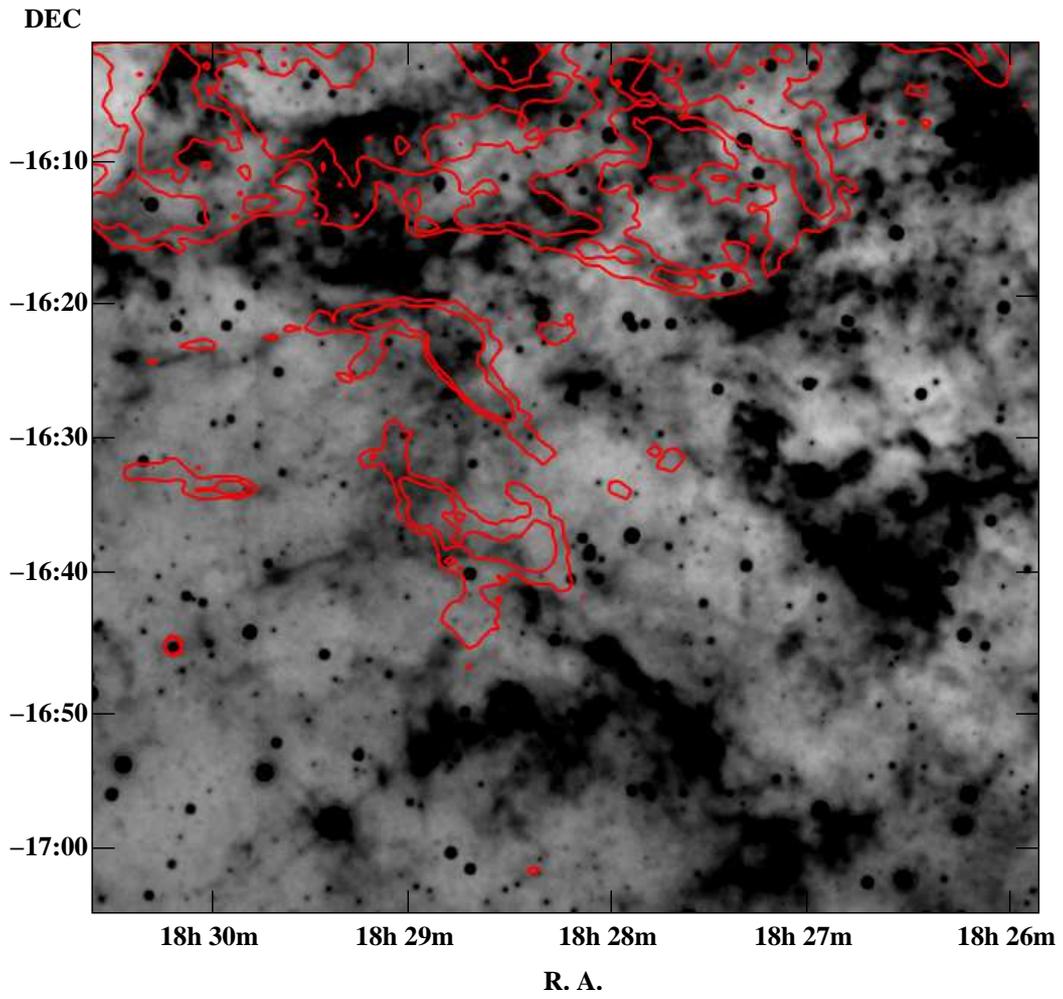}
\caption{The correlation between the infrared WISE emission
at  22$\mu$m image and the \hnii\  emission (red lines). The red contours
 values are 70 (low) and 130 (high) $\times$\flux.}
\label{fig6}
\end{figure*}

The SNR origin of the proposed candidate remnants is strongly suggested
by the positions of the line ratios in Fig.~\ref{fig4} compared with
those of Herbig--Haro objects, H{\sc ii} regions and
planetary nebulae (PNe). They follow closely the shape of those
observed for those of shock ionized evolved SNRs.  

\par 
The \ha/\hbeta\ ratios in Table \ref{table4} can be used to estimate
the variations in logarithmic extinction coefficient c over these
sources, assuming an intrinsic ratio of 3 and the interstellar
extinction curve as implemented in the nebular
package \citep{sha95} within the IRAF software. 
An interstellar extinction c (see Table~\ref{table4}) between 0.2 and 1.1 or an A$_{\rm V}$~between 0.4 and 2.2 were measured, respectively. 
We have also determined the electron density measuring the
density--sensitive line ratio of \siirat. The measured densities lie
below 240 \dens. %
\par
The candidate remnants under
investigation have not been studied in the past hence the current stage
of their evolution is unknown. Our aim is to provide a first indication of their stage of evolution by estimating basic SNR parameters, assuming that the temperature is close to 10$^{4}$ K. 

Estimated values of N$_{\rm H}$ between 4.8 and 6.8 $\times 10^{21}$~cm$^{-2}$ and N$_{\rm H}$ between 3.2 and 7.4 $\times 10^{21}$~cm$^{-2}$ are given by \citet{dic90} and \citet{kal05}
respectively, for the column density in the direction of the candidate remnants. 
Using the relations of \citet{ryt75} and \citet{pre95}, we obtain N$_{\rm H}$ between 0.9 and 4.9 $\times 10^{21}~{\rm cm}^{-2}$~for the minimum and maximum c values calculated
from our spectra.  In Table~\ref{table3}, we present the estimated values of N$_{\rm H}$ for each candidate remnant, where it can be seen that the values based on the optical data and the statistical relations are consistent with the estimated galactic N$_{\rm H}$~ from \citet{kal05} and less by that estimated from \citet{dic90}. However, it should be noted that the slightly higher values calculated by the latter method can be explained by the fact that it also covers gas beyond the area of interest.
Assuming that they are still in the
adiabatic phase of their evolution the preshock cloud density n$_{\rm
c}$ can be measured by using the relationship \citep{dop79}

\begin{equation}
{\rm n_{[SII]} \simeq\ 45\ n_c V_{\rm s}^2}~{\rm cm^{-3}},
\end{equation}

where ${\rm n_{[SII]}}$ is the electron density derived from the
sulfur line ratio and V$_{\rm s}$ is the shock velocity into the
clouds in units of 100 \vel. Furthermore, the blast wave energy can be
expressed in terms of the cloud parameters by using the equation given
by \citet{mck75}

\begin{equation}
{\rm E_{51}} = 2 \times 10^{-5} \beta^{-1} 
{\rm n_c}\ V_{\rm s}^2 \ 
{\rm r_{s}}^3 \ \ {\rm erg}.
\end{equation}

The factor $\beta$ is approximately equal to 1 at the blast wave
shock, ${\rm E_{51}}$ is the explosion energy in units of 10$^{51}$
erg and {\rm r$_{\rm s}$} the radius of the remnant in pc. 

By using the upper limit on the electron density of 240 \dens, which was
derived from our spectra, we obtain from Eq. (1) that ${\rm n_c}
V_{\rm s}^2 < 5.3$. Then Eq. (2) becomes ${\rm E_{51}} < \alpha \times
10^{-3}~{\rm D_{1 kpc}^3}$, where $\alpha$ is a value depends on the diameter of each candidate remnant and ${\rm D_{1 kpc}}$~the distance to
the remnant in units of 1 kpc. For the different diameters of the remnants and assuming the typical value of 1 for the supernova
explosion energy (E$_{51}$), we derive that the candidate SNRs lie at distances greater than 8 kpc. Since, there are no other measurements of the interstellar density
n$_{0}$, values of 0.1 and 1.0 will be examined. 
Then, the lower interstellar density
of $\sim$0.1 cm$^{-3}$~suggests that their distance is between 10 and 22 kpc while for n$_{0} \approx 1~{\rm cm}^{-3}$~it is between 1.0 and 2.2 kpc,
for the lower and higher column densities calculated above.
Combining the previous results, values between 0.1 and 0.2 cm$^{-3}$~for the interstellar density seem to be more probable. It should also be noted that the ambient density of gas around the candidate SNRs is not the same so this might also change the estimated vaues.

We also searched for pulsars in the region using the ATNF Pulsar Catalogue \citep{man05}. In total, we found 8 pulsars within a 1.5$\degr$ diameter circle away from the center of each candidate SNR. In Table 5, we present their names, coordinates and rotation period as well as which candidate SNRs fulfill the 1.5$\degr$ limit. The closest one is  PSR1826-1526 to G 15.8$-$1.9 at a distance of 0.7$\degr$. It is not clear at the moment if any of these pulsars are related to the candidate SNRs, however, the existence of a significant number of pulsars very close to the area of the candidate SNRs it is another strong indication of the existence of more than one SNRs in the region. It is more plausible that due to their distance from the candidate SNRs they are not associated with them since the closest pulsar is about 4 radii from the nearest SNR or $>$ 75 pc in the plane of the sky at a distance of 8 kpc.  However, if they are not in the plane of the sky which is probably the case, then these numbers might change. Also, they might be at different distances which change the numbers, too. Therefore, their correlation cannot be confirmed or ruled out and it should be further examined in the future in detail.

It is possible that this irregular group of
filaments is part of a wider structure but is being seen through holes in
intervening clouds, leading to patchy optical interstellar
extinction. The current data are not sufficient to claim strong a
correlation.
Furthermore, there is a possibility that the detected optical emission could be part of a number of supernova explosions in the area.
However, since neither the distance nor the interstellar
medium density are accurately known, we cannot confidently determine
the current stage of evolution of the candidate remnants and  more observations are needed. 

Further study of the area  would benefit from higher resolution multiwavelength observations (optical, radio and X--ray) and will help to clarify the current uncertainties. In particular, higher resolution imaging observations will verify their filamentary structure and confirm their morphological appearance, while kinematic observations will help to determine their 3-D morphology and measure expansion velocities. X--ray observations will help in order to clarify if there is any correlation of the faint and diffuse X--ray emission with the optical filaments and provide more information about their evolutionary stage, while radio observations would also be useful to examine their non--thermal spectral index and confirm their SNR nature.
\begin{acknowledgements}
We thank the referee for his constructive comments and suggestions which helped to improve the manuscript significantly. Skinakas Observatory is a collaborative project of the University of
Crete, the Foundation for Research and Technology-Hellas and the
Max-Planck-Institut f\"ur Extraterrestrische Physik. This research
made use of data from SuperCOSMOS \ha\ Survey (AAO/UKST), from the ATNF Pulsar Catalogue and from the NASA/IPAC Infrared Science Archive.
\end{acknowledgements}

\newpage

%
\begin{table*}  
\caption[]{Imaging and Spectral log}  
\label{table1}
\begin{tabular}{lcclc}  
\noalign{\smallskip}  
\hline  
\multicolumn{5}{c}{IMAGING} \\  
\hline
Filter & $\lambda_{\rm c}$ & $\Delta \lambda$ & Total exp. time &
telescope \\
  & (\AA) & (\AA) & (sec)&  \\
\hline
\HNII & 6570 & 75 & 4800 (2)\tablefootmark{a} & 0.3m Skinakas\\
\OIII & 5005 & 28 & 4800 (2)  & 0.3m Skinakas\\
\SII & 6720 & 18 & 4800 (2)  & 0.3m Skinakas\\
Cont blue & 5470 & 230 & 180 & 0.3m Skinakas\\
Cont red & 6096 & 134 & 180  & 0.3m Skinakas\\
\HNII & & 80 & 10800 & 1.2m AAO/UKST\\
\hline
\multicolumn{5}{c}{SPECTROSCOPY} \\  
\hline  
Slit position & \multicolumn{2}{c}{Slit centers} & Offset\tablefootmark{b} & Aperture
length\tablefootmark{c} \\
 & $\alpha$ & $\delta$ &  & \\
 & (h m s) & (\degr\ \arcmin\ \arcsec) & (arcsec) & (arcsec) \\  
\hline
1a & 18 29 09 & $-$16 31 32 & 2.4 N & 10.6 \\
1b & 18 29 09 & $-$16 31 32 & 28.9 S & 18.9 \\
2a & 18 28 58 & $-$16 36 00 & 95.0 N & 26.0 \\
2b & 18 28 58 & $-$16 36 00 & 70.2 N & 11.8 \\
2c & 18 28 58 & $-$16 36 00 & 19.5 N & 13.0 \\
3a & 18 28 49 & $-$16 24 55 & 137.5 N & 27.1\\
3b & 18 28 49 & $-$16 24 55 & 33.6 N & 43.6\\
3c & 18 28 49 & $-$16 24 55 & 20.1 S & 23.6\\
4  & 18 29 55 & $-$16 34 12 & 16.6 S & 30.7\\
5  & 18 28 42 & $-$16 15 02 & 37.2 S & 60.2\\
6a & 18 27 41 & $-$16 17 39 & 74.3 S & 30.7\\
6b & 18 27 41 & $-$16 17 39 & 116.2 S & 20.0\\
7  & 18 27 26 & $-$16 07 50 & 32.4 N & 31.9\\
8  & 18 26 12 & $-$16 01 58 & 27.7 N & 23.6\\
\hline  
\end{tabular}
\tablefoot{
\tablefoottext{a}{Numbers in parentheses represent the number of individual
frames.}\\
\tablefoottext{b}{Spatial offset from the slit center in arcsec: N($=$North),
S($=$South).}\\ 
\tablefoottext{c}{Aperture lengths for each area in arcsec.}\\}
\end{table*}  

\begin{table*}
\caption[]{Typically fluxes measured over the brightest filaments.}
\label{table2}
\begin{tabular}{lllllllll}
\hline
\noalign{\smallskip}
 & Slit 1 & Slit 2 & Slit 3 & Slit 4 & Slit 5 & Slit 6 & Slit 7 & Slit 8 \\
\hline
\hnii\  & 42.8 & 62.8 & 114.3 & 64.1 & 67.2 & 50.4 & 57.9 & 53.8 \\
\hline
\sii\   & 19.5 & 18.8 & 40.0 & 16.6 & 15.8 & 15.6 & 17.1 & 13.5 \\
\hline
\oiii\ & 236 & \multicolumn{7}{c}{$<$45\tablefootmark{a}} \\
\hline
{\bf \sii/\ha\tablefootmark{b}} & 0.91 & 0.60 & 0.70 & 0.52 & 0.47 & 0.62 & 0.59 & 0.50 \\
\hline
\end{tabular}
\tablefoot{
Fluxes in units of \flux. Median values over a 40\arcsec $\times$ 40\arcsec\ box from selected areas on the bright filaments. \\
\tablefoottext{a}{3$\sigma$~upper limit.}\\
\tablefoottext{b}{The \nitrogen\ contribution has been removed by using the spectroscopic results}}
 \end{table*}

\begin{table*}
\caption[]{New candidate SNRs.}
\label{table3}
\begin{tabular}{ccccccc}
\hline
\noalign{\smallskip}
SNR name & \multicolumn{2}{c}{SNR center} & Diameter & \multicolumn{2}{c}{N$_{\rm H}$\tablefootmark{a}} & {N$_{\rm H}$\tablefootmark{b}}  \\
 & $\alpha$ & $\delta$ &   & Kal\tablefootmark{c} & D\&L\tablefootmark{d} & Ral\tablefootmark{e}  \& P\&S\tablefootmark{f} \\
 & (h m s) & (\degr\ \arcmin\ \arcsec) & (arcmin) & ($\times 10^{21}$ cm$^{-2}$) & ($\times 10^{21}$ cm$^{-2}$) & ($\times 10^{21}$ cm$^{-2}$) \\  
\hline
G 15.6$-$2.6 & 18 28 46.1 & $-$16 30 40 & 3.7$\times$10.2 & 4.6 & 6.8  & 1.0--4.9 \\
G 15.8$-$2.8 & 18 29 35.8 & $-$16 27 40 & 5.1$\times$12.9 & 3.2 & 4.8 & 1.0-4.9 \\
G 15.8$-$2.2 & 18 27 38.8 & $-$16 12 17 & 6.5$\times$8.7 & 4.6 & 6.8 & 2.6--3.3 \\
G 15.8$-$1.9 & 18 26 30.7 & $-$16 08 00 & 7.7$\times$8.3  & 7.4 & 6.8 & 0.7--4.8 \\
G 16.2$-$2.5 & 18 29 23.5 & $-$16 03 06 & 3.0$\times$7.2 & 4.6 & 6.2  & 2.8--3.5 \\
G 15.6$-$2.7 & 18 28 53.5 & $-$16 35 34 & 1.1$\times$1.1 & 4.6 & 6.8 & 2.0-4.8 \\
\hline
\end{tabular}
\tablefoot{
\tablefoottext{a}{N$_{\rm H}$ derived by the statistical relations in the direction of the candidate SNRs.\\}
\tablefoottext{b}{N$_{\rm H}$ calculated using current observations (min and max E(B--V) taken from Table 4) and the statistical relations.\\}
\tablefoottext{c}{\citet{kal05}}; \tablefoottext{d}{ \citet{dic90}};
\tablefoottext{e}{\citet{ryt75}}; \tablefoottext{f}{\citet{pre95}}. \\}
\end{table*}
\begin{table*}
\caption[]{Relative line fluxes.}
\label{table4}
\begin{tabular}{llllllllllllllll}
\hline
\noalign{\smallskip}
 & \multicolumn{3}{c}{Slit 1a} & \multicolumn{3}{c}{Slit 1b} 
& \multicolumn{3}{c}{Slit 2a} & \multicolumn{3}{c}{Slit 2b} &
\multicolumn{3}{c}{Slit 2c} \\ 
Line (\AA) & F\tablefootmark{a} & I \tablefootmark{b} & S/N\tablefootmark{c} & F & I & S/N
& F & I & S/N & F & I & S/N & F & I & S/N \\
\hline
\hbeta\ 4861 & 23 & 35 & 13 & 15 & 35 & 2 & 19 & 35
& 27 & 21 & 35 & 18 & 24 & 35 & 21 \\
\oxygen\ 4959 & 13 & 20 & 7 & 66 & 147 & 5 &  $-$ &
$-$ & $-$ & $-$ & $-$ & $-$ & $-$ & $-$ & $-$ \\
\oxygen\ 5007 & 36 & 53 & 23 & 187 & 405 & 13 & 
$-$ & $-$ & $-$ & $-$ & $-$ & $-$ & $-$ & $-$ & $-$ \\
\oi\ 6300 & 14 & 14 & 17 & $-$ & $-$ & $-$ & $-$ & $-$ & $-$ & $-$ &
$-$ & $-$ & $-$ & $-$ & $-$ \\
\nitrogen\ 6548 & 32 & 32 & 36 & 36 & 36 & 5 & 14
& 14 & 42 & 15 & 15 & 22 & 13 & 12 & 42 \\
\ha\ 6563 & 100 & 100 & 89 & 100 & 100 & 14 &  100
& 100 & 216 & 100 & 100 & 123 & 100 & 100 & 137 \\
\nitrogen\ 6584 & 95 & 95 & 82 & 100 & 99 & 14 & 40 & 39 & 103 & 44 &
43 & 58 & 42 & 42 & 66 \\
\sulfur\ 6716 & 63 & 61 & 59 & 74 & 69 & 10 & 
27 & 26 & 71 & 31 & 29 & 44 & 30 & 29 & 44 \\
\sulfur\ 6731 & 49 & 47 & 47 & 53 & 49 & 7 &  22
& 21 & 56 & 26 & 25 & 38 & 26 & 23 & 38 \\
\hline
Absolute \ha\ flux\tablefootmark{d} & \multicolumn{3}{c}{7.5} &
\multicolumn{3}{c}{0.62} & \multicolumn{3}{c}{16.7} &
\multicolumn{3}{c}{13.3} & \multicolumn{3}{c}{11.9} \\
\sulfur/\ha\ & \multicolumn{3}{c}{1.08 $\pm$ 0.02} &
\multicolumn{3}{c}{1.20$\pm$ 0.1} & \multicolumn{3}{c}{0.47$\pm$ 0.03}
& \multicolumn{3}{c}{0.54 $\pm$0.05} & \multicolumn{3}{c}{0.51$\pm$
0.04} \\
F(6716)/F(6731) & \multicolumn{3}{c}{1.30 $\pm$ 0.04} &
\multicolumn{3}{c}{1.4$\pm$ 0.3} & \multicolumn{3}{c}{1.3$\pm$ 0.1} &
\multicolumn{3}{c}{1.2$\pm$ 0.1} &\multicolumn{3}{c}{1.3$\pm$ 0.1} \\
\nitrogen/\ha\ & \multicolumn{3}{c}{1.28$\pm$ 0.02} &
\multicolumn{3}{c}{1.35$\pm$ 0.1} & \multicolumn{3}{c}{0.53$\pm$0.03}
& \multicolumn{3}{c}{0.58$\pm$ 0.06} & \multicolumn{3}{c}{0.54$\pm$
0.05} \\
c(\hbeta)\tablefootmark{e} & \multicolumn{3}{c}{0.52$\pm$ 0.10} &
\multicolumn{3}{c}{1.1$\pm$ 0.6} & \multicolumn{3}{c}{0.74$\pm$ 0.05}
& \multicolumn{3}{c}{0.64$\pm$ 0.07} & \multicolumn{3}{c}{0.50$\pm$
0.06} \\
E$_{\rm B-V}$ & \multicolumn{3}{c}{0.37 $\pm$ 0.07} &
\multicolumn{3}{c}{0.7 $\pm$ 0.4}& \multicolumn{3}{c}{0.51 $\pm$ 0.03}
& \multicolumn{3}{c}{0.44 $\pm$ 0.05} & \multicolumn{3}{c}{0.34 $\pm$
0.04}\\
\hline
 & \multicolumn{3}{c}{Slit 3a} &
\multicolumn{3}{c}{Slit 3b} & \multicolumn{3}{c}{Slit 3c} &
\multicolumn{3}{c}{Slit 4} & \multicolumn{3}{c}{Slit 5} \\ 
Line (\AA) & F & I & S/N & F & I & S/N & F & I & S/N & F & I & S/N & F
& I & S/N \\
\hline 
\hbeta\ 4861 & 15 & 35 & 18 & 20 & 35 & 44 & 18
& 35 & 32 & 28 & 35 & 16 & 20 & 35 & 21 \\
\oi\ 6300  & 9 & 10 & 26 & 5 & 5 & 33 & 7 & 7 & 31 &
 $-$ & $-$ & $-$ & 24 & 25 & 61 \\
\oi\ 6363 & $-$ & $-$ & $-$ & $-$ & $-$ & $-$ & $-$ &
$-$ & $-$ & $-$ & $-$ & $-$ & 12 & 12 & 27 \\
\nitrogen\ 6548 & 17 & 16 & 38 & 16 & 16 & 85 & 18 & 18
& 66 & 15 & 14 & 21 & 15 & 15 & 37 \\
\ha\ 6563  & 100 & 100 & 165 & 100 & 100 & 337 & 100
& 100 & 243 & 100 & 100 & 115 & 100 & 100 & 194 \\
\nitrogen\ 6584 & 96 & 95 & 73 & 44 & 44 & 180 & 49 &
48 & 141 & 47 & 47 & 58 & 57 & 56 & 122 \\
\sulfur\ 6716 & 63 & 60 & 28 & 28 & 26 & 115 & 31 & 29
& 94 & 27 & 26 & 36 & 27 & 26 & 57 \\
\sulfur\ 6731 & 49 & 45 & 16 & 20 & 18 & 88 & 21 & 19 &
70 & 21 & 20 & 28 & 20 & 19 & 42 \\
\hline 
Absolute \ha\ flux & \multicolumn{3}{c}{14.1} &
\multicolumn{3}{c}{27.0} & \multicolumn{3}{c}{26.3} &
\multicolumn{3}{c}{10.7} & \multicolumn{3}{c}{20.2} \\
\sulfur/\ha\ & \multicolumn{3}{c}{1.05$\pm$ 0.09} &
\multicolumn{3}{c}{0.45 $\pm$0.02} & \multicolumn{3}{c}{0.49 $\pm$
0.03} & \multicolumn{3}{c}{0.47$\pm$ 0.05} & \multicolumn{3}{c}{0.45
$\pm$0.04} \\
F(6716)/F(6731) & \multicolumn{3}{c}{1.3$\pm$ 0.2} &
\multicolumn{3}{c}{1.41$\pm$ 0.07} & \multicolumn{3}{c}{1.5 $\pm$ 0.1}
& \multicolumn{3}{c}{1.3$\pm$ 0.2} & \multicolumn{3}{c}{1.4$\pm$ 0.1} \\
\nitrogen/\ha\ &\multicolumn{3}{c}{1.1$\pm$0.1} &
\multicolumn{3}{c}{0.60$\pm$ 0.02} & \multicolumn{3}{c}{0.66$\pm$
0.04}& \multicolumn{3}{c}{0.61$\pm$ 0.07} &
\multicolumn{3}{c}{0.71$\pm$ 0.06} \\
c(\hbeta) & \multicolumn{3}{c}{1.04$\pm$ 0.07} &
\multicolumn{3}{c}{0.70$\pm$ 0.03}& \multicolumn{3}{c}{0.82$\pm$
0.04}& \multicolumn{3}{c}{0.27$\pm$ 0.08} &
\multicolumn{3}{c}{0.69$\pm$ 0.06} \\
E$_{\rm B-V}$ & \multicolumn{3}{c}{0.57 $\pm$ 0.02}
&\multicolumn{3}{c}{0.19 $\pm$ 0.05}& \multicolumn{3}{c}{0.72 $\pm$
0.05} & \multicolumn{3}{c}{0.48 $\pm$ 0.02} & \multicolumn{3}{c}{0.48
$\pm$ 0.04} \\
\hline
\end{tabular}
\end{table*}
\longtab{4}{
\begin{table*}
\caption[]{Continued.}
\label{table4}
\begin{tabular}{llllllllllllllll}
\hline
\noalign{\smallskip}&\multicolumn{3}{c}{Slit 6a} &
\multicolumn{3}{c}{Slit 6b} & \multicolumn{3}{c}{Slit 7} &
\multicolumn{3}{c}{Slit 8} & \multicolumn{3}{c}{} \\
Line (\AA) & F & I & S/N & F & I & S/N & F & I & S/N & F & I & S/N &
\multicolumn{3}{c}{} \\
\hline
\hbeta\ 4861  & 27 & 35 & 13 & 30 & 35 & 24 & 16 & 35 &
8 & 19 & 35 & 10 & \multicolumn{3}{c}{} \\
\oi\ 6300  & 31 & 31 & 35 & 21 & 21 & 40 & $-$ & $-$ &
$-$ & $-$ & $-$ & $-$ & \multicolumn{3}{c}{} \\
\nitrogen\ 6548  & 37 & 37 & 36 & 20 & 21 & 39 & 23 & 23
& 22 & 16 & 16 & 23 & \multicolumn{3}{c}{} \\
\ha\ 6563 & 100 & 100 & 84 & 100 & 100 & 146 & 100 &
100 & 81 & 100 & 100 & 111 & \multicolumn{3}{c}{} \\
\nitrogen\ 6584 & 95 & 94 & 73 & 66 & 65 & 96 & 67 &
66 & 55 & 62 & 61 & 71 & \multicolumn{3}{c}{} \\
\sulfur\ 6716 & 45 & 44 & 39 & 26 & 25 & 41 & 31 & 29 &
27 & 29 & 28 & 38 & \multicolumn{3}{c}{} \\
\sulfur\ 6731 & 30 & 29 & 27 & 19 & 19 & 30 & 21 & 19 &
20 & 21 & 20 & 30 & \multicolumn{3}{c}{} \\
\hline
Absolute \ha\ flux & \multicolumn{3}{c}{8.3} &
\multicolumn{3}{c}{13.0} & \multicolumn{3}{c}{19.7} &
\multicolumn{3}{c}{41.3} & \multicolumn{3}{c}{} \\
\sulfur/\ha\ & \multicolumn{3}{c}{0.7 $\pm$ 0.1} &
\multicolumn{3}{c}{0.44$\pm$ 0.03} & \multicolumn{3}{c}{0.5$\pm$ 0.1}
& \multicolumn{3}{c}{0.48 $\pm$0.08} & \multicolumn{3}{c}{} \\
F(6716)/F(6731) & \multicolumn{3}{c}{1.5 $\pm$ 0.2} &
\multicolumn{3}{c}{1.3$\pm$ 0.1} & \multicolumn{3}{c}{1.5$\pm$ 0.4}&
\multicolumn{3}{c}{1.4$\pm$ 0.3} & \multicolumn{3}{c}{} \\
\nitrogen/\ha\ & \multicolumn{3}{c}{1.3$\pm$ 0.2} &
\multicolumn{3}{c}{0.86$\pm$ 0.07} & \multicolumn{3}{c}{0.9$\pm$ 0.2}
& \multicolumn{3}{c}{0.8$\pm$ 0.1} & \multicolumn{3}{c}{} \\
c(\hbeta) & \multicolumn{3}{c}{0.34$\pm$ 0.09} &
\multicolumn{3}{c}{0.20$\pm$ 0.05} & \multicolumn{3}{c}{1.0$\pm$ 0.2}&
\multicolumn{3}{c}{0.7$\pm$ 0.1} & \multicolumn{3}{c}{} \\
E$_{\rm B-V}$ & \multicolumn{3}{c}{0.23 $\pm$ 0.07} &
\multicolumn{3}{c}{0.13 $\pm$ 0.04}&\multicolumn{3}{c}{0.7 $\pm$ 0.1}
& \multicolumn{3}{c}{0.51 $\pm$ 0.09} & \multicolumn{3}{c}{} \\
\hline
\end{tabular}
\tablefoot{The errors of the emission line ratios are calculated through standard error propagation.\\
\tablefoottext{a}{Observed fluxes normalized to F(H$\alpha$)=100 and
uncorrected for interstellar extinction. \\}
\tablefoottext{b}{Intrinsic surface brightness normalized to I(H$\alpha$)=100
and corrected for interstellar extinction. \\}
\tablefoottext{c}{Numbers represent the signal to noise ratio of the quoted
fluxes.\\}
\tablefoottext{d}{In units of \flux.\\}
\tablefoottext{e}{The logarithmic extinction is derived by c =
1/0.348$\times$log((\ha/\hbeta)$_{\rm obs}$/2.85).\\}}
\end{table*}}
\begin{table*}
\caption[]{Pulsars found within a region of 1.5\degr\ from the candidate SNRs.}
\label{table5}
\begin{tabular}{cccccc}
\hline
\noalign{\smallskip}
Pulsar name & \multicolumn{2}{c}{Pulsar center} & Period &  candidate SNR\tablefootmark{a} & References\\
 & $\alpha$ & $\delta$ &   &  & \\
 & (h m s) & (\degr\ \arcmin\ \arcsec) & (s) &  &  \\  
\hline
J1822$-$1606 & 18 22 23.0 & $-$16 05 59.0 & 8.4377 & 3, 4  & (1),(2) \\
J1822$-$1617 & 18 22 36.6 & $-$16 17 35.0 & 0.8311 & 3, 4 & (3) \\
J1823$-$1526 & 18 23 21.4 & $-$15 26 22.0 & 1.6254 & 3, 4 & (3)  \\
J1824$-$1500 & 18 24 14.1 & $-$15 00 33.0 & 0.4122 & 3, 4 & (3) \\
J1825$-$1446 & 18 25 02.9 & $-$14 46 52.6 & 0.2792 & 4 & (4),(5)\\
J1826$-$1526 & 18 26 12.6 & $-$15 26 03.0 & 0.3820 & 1-6 & (6) \\
J1829$-$1751 & 18 29 43.1 & $-$17 51 03.9 & 0.3071 & 1, 2, 6 & (5),(7),(8)\\
J1834$-$1710 & 18 34 53.4 & $-$17 51 03.9 & 0.3583 & 2 & (9)\\
\hline
\end{tabular}
\tablefoot{
\tablefoottext{a}{ Candidate SNRs numbered according to their order presented in Table~\ref{table3}.}\\
(1)  \citet{rea11}; (2) \citet{cum11}; (3) \citet{hob04a}; (4) \citet{cli86}; \\
(5) \citet{hob04b}; (6) \citet{mor02}; (7) \citet{dav72}; (8) \citet{zou05}; (9) \citet{kra03}.
}
\end{table*}

\end{document}